\documentclass[twocolumn,twoside,slac_two]{revtex4}
\usepackage{graphicx}
\usepackage{fancyhdr}
\begin{document}

\title{AGILE Observations of Terrestrial Gamma-Ray Flashes}

%

\author{M. Marisaldi, F. Fuschino, C. Labanti, A. Bulgarelli, F. Gianotti, M. Trifoglio}
\affiliation{INAF IASF Bologna, Via Gobetti 101, 40129 Bologna, Italy}
\author{M. Tavani, A. Argan, E. Del Monte}
\affiliation{INAF IASF Roma, Via del Fosso del Cavaliere 100, I-00133 Roma, Italy}
\author{F. Longo, G. Barbiellini}
\affiliation{Dipartimento di Fisica Universit\`a di Trieste and INFN Trieste, Via Valerio 2, I-34127 Trieste, Italy}
\author{A. Giuliani}
\affiliation{INAF IASF Milano, Via E. Bassini 15, I-20133 Milano, Italy}
\author{A. Trois}
\affiliation{INAF Osservatorio Astronomico di Cagliari, loc. Poggio dei Pini, strada 54, 09012 Capoterra (CA), Italy}
\author{on behalf of the AGILE team}

\begin{abstract}
The AGILE satellite, operating since mid 2007 and primarily devoted to
high-energy astrophysics, is one of the only three currently operating
space instruments capable of detecting Terrestrial Gamma-Ray Flashes
(TGFs), together with RHESSI and $Fermi$-GBM. Thanks to the AGILE
Mini-Calorimeter instrument energy range extended up to 100MeV and its
flexible trigger logic on sub-millisecond time scales, AGILE is
detecting more than 10 TGFs/month, adding a wealth of observations
which pose severe constrains on production models. The main AGILE
discoveries in TGF science during two and a half years of observations
are the following: 1) the TGF spectrum extends well above 40 MeV, 2)
the high energy tail of the TGF spectrum is harder than expected and
cannot be easily explained by previous theoretical models, 3) TGFs can be localized from space using high-energy photons detected by the AGILE gamma-ray imaging detector. In this presentation we will describe the characteristics of the 2.5-years AGILE TGF sample, focusing on the recent results concerning the TGF high-energy spectral characteristics.
\end{abstract}

\maketitle

\thispagestyle{fancy}


\section{INTRODUCTION}

Terrestrial Gamma-Ray Flashes (TGFs) are very  short 
(lasting  up to a few milliseconds) bursts of high-energy  photons above
100 keV, first  detected by the BATSE instrument on board the
Compton Observatory \cite{Fishman1994}, and later extensively observed by the RHESSI satellite \cite{Smith2005,Grefenstette2009}.   
TGFs have been associated with strong thunderstorms mostly concentrated in the Earth's equatorial and tropical regions, at a typical altitude of 15--20~km \cite{Dwyer2005}. 
TGFs are widely believed to be produced by Bremsstrahlung in the atmospheric layers by a population of runaway electrons accelerated to relativistic energies by strong electric fields inside or above thunderclouds. The secondaries generated during the acceleration process can be accelerated as well driving an avalanche multiplication \cite{Gurevich1992}, commonly referred to as Relativistic Runaway Electron Avalanche (RREA), possibly further enhanced by means of a relativistic feedback mechanism \cite{Dwyer2007,Dwyer2008}. 
Although being widely accepted as the underlying physical process in TGF production, the RREA mechanism alone is not sufficient to explain the rich phenomenology of TGFs, especially the observed fluence, and there is no consensus yet on the underlying physical conditions, production sites, radiation efficiencies and maximal energies. 

Recently, the AGILE satellite added a wealth of spectral and geographical data on TGFs and established itself as a major player in TGF observation, together with the RHESSI \cite{Smith2005} and $Fermi$-GBM detectors \cite{Briggs2010}.

AGILE \cite{Tavani2008b} is a mission of the Italian Space Agency
(ASI) dedicated to astrophysics in the gamma-ray energy range
30~MeV -- 30~GeV, with a monitor in the X-ray band 18~keV --
60~keV \cite{Feroci2007}, operating since April 2007 in a low inclination
($2.5^\circ$) equatorial Low-Earth Orbit at 540~km altitude. The AGILE Gamma-Ray Imaging Detector (GRID) is a pair-tracking telescope
based on a tungsten-silicon tracker \cite{Prest2003}, sensitive in the 30~MeV -- 30~GeV energy range. The imaging
principle is based on the reconstruction of the tracks left in the silicon
detection planes by the electron-positron pairs produced by
the primary photon  converting mainly in the tracker  tungsten planes. 
A Mini-Calorimeter
(MCAL) \cite{Labanti2009}, based on CsI(Tl) scintillating bars for
the detection of gamma-rays in the range 300~keV -- 100~MeV,
 and a plastic anti-coincidence detector \cite{Perotti2006}
complete the high-energy instrument. MCAL can work also as an
independent gamma-ray transient detector with a dedicated
 trigger logic acting on several time scales
spanning four orders of magnitude between 290~$\mu$s and 8~seconds
\cite{Fuschino2008,Argan2004}. 

Thanks to its flexible trigger logic on  sub-millisecond time scales \cite{Fuschino2008}, MCAL proved to be a very efficient instrument for TGF detection. 
In this paper we will review the AGILE achievements in TGF science,
focusing on our results on TGF high-energy spectral characteristics
and on the correlation with global lightning activity.

\section{AGILE RESULTS IN TGF SCIENCE}

The average MCAL detection rate is $\sim10$~TGFs/month, with the
current severe selection criteria based on hardness ratio and fluence
\cite{Marisaldi2010}.
When a trigger is issued by the onboard logic, data are collected on
an event-by-event basis so that, for every photon, energy,
timing with $2\mu s$ accuracy and topological informations (i.e. the
fired detector address) are saved and sent to ground.
For a trigger to be classified as a valid TGF candidate, at least 10 photons and a hardness ratio $HR \ge 0.5$ are required, where $HR$ is defined as the ratio between the number of counts with energy greater than 1.4~MeV and the number of counts with energy lower than 1.4~MeV. 
There is evidence that relaxed selection criteria may be applied,
resulting in the detection of a fainter/softer population which can
increase the detection rate more than 50\% (Marisaldi, M. \textit{et
  al.}, in preparation).

The main AGILE discoveries in TGF science during two and a half years of observations are the following:
\begin{itemize}
\item[-] the TGF spectrum extends at least up to 40~MeV \cite{Marisaldi2010} (well above the previous 20~MeV limit set by RHESSI  \cite{Smith2005});
\item[-] the high energy tail of the TGF spectrum is harder than expected and cannot be easily explained by previous theoretical models \cite{Tavani2010};
\item[-] TGF can be localized from space using high-energy photons detected by the AGILE silicon tracker \cite{Marisaldi2010b}
\item[-] TGFs are not a random sub-sample of global lightning activity as detected from space \cite{Fuschino2011}. Moreover, significant regional differences exist, both in the degree of correlation and in the TGF/flash ratio. 
\end{itemize}

\subsection{TGFs high-energy spectrum}

The cumulative spectrum of the first 34 high-confidence TGFs detected by AGILE shows
a good agreement with the RHESSI cumulative spectral shape below
20~MeV, when the detector response matrices for both instruments are
taken into account, but demonstrates also that events of energy up to
40~MeV can be observed \cite{Marisaldi2010}. These early observations
substantially doubled the previously known TGF energy range set by
RHESSI \cite{Smith2005}, as later confirmed by the observations of
$Fermi$-GBM  \cite{Briggs2010}.

\begin{figure}
\includegraphics{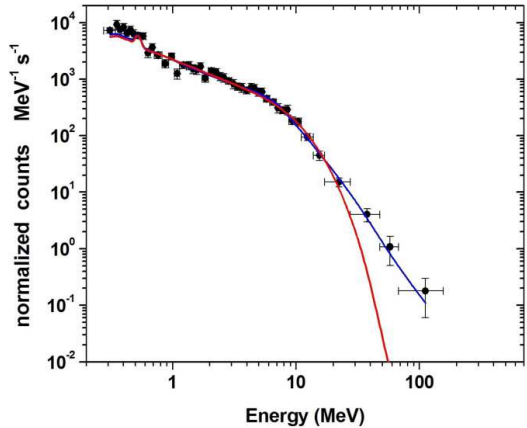}
 \caption{
  The background subtracted cumulative counts spectrum of the 130 TGFs detected
  by AGILE-MCAL during the period June, 2008 - January, 2010. The blue line shows the
  broken PL fit (see text), and the red curve is a pre-AGILE phenomenological model
  $ F(E) \sim E^{-\alpha} \, e^{(-E/E_c)}$,
  of  index $\alpha  = 0.4 \pm 0.2$, and exponential cutoff energy $
  E_c =  6.6 \pm 1.2$~MeV. Figure from \cite{Tavani2010}.}
\label{fig-3}
\end{figure}

The high energy spectral shape of TGFs became even more puzzling when
increased statistics was achieved. Fig. \ref{fig-3} shows the background-subtracted cumulative energy spectrum
of a TGF sample of 130 events, satisfying the same stringent selection criteria,
detected during the period June 2008 - January 2010. The relevant
cumulative background as a function of energy is calculated for
events detected in the time interval $T_{o}$+1 sec - $T_{o}$+21 sec,
where $T_{o}$ is the TGF start time. This method takes into
account  the 20$\%$ orbital modulation of the MCAL background in a
satisfactory way. We note that the MCAL background level on TGF
timescales is quite low (because of the MCAL anticoincidence vetoing, and
of the satellite equatorial orbit), being 0.35 events/ms on average.
The MCAL detector response for different off-axis
angles has been derived by a combination of simulations and
calibration data obtained up to a few MeV with radioactive
sources, and up to 460 MeV at the beam test facility of the
National Laboratories of Frascati (Italy). In-orbit calibration
consistency checks were obtained for different Gamma-Ray Bursts at
different angles. 
Remarkably, we find that the TGF spectrum
extends up to 100 MeV with no exponential attenuation. 
Our data show the existence of a high-energy spectral component in addition
to the well-known power-law (PL) component extending up to $\sim$
10 MeV. The additional component constitutes $\sim$10$\%$ of the
total emitted energy. A broken PL fit of the two components  gives
a differential photon energy flux F(E) $\sim$ E $^{-0.5 \pm 0.1}$
for 1 MeV $<$E$<$ E$_{c}$, and $F(E) \sim E ^{-2.7 \pm 0.1}$  for
$E_{c} < E
 < $~100 MeV, with $E_{c} = (7.1 \pm 0.5)$~MeV (all quoted parameter errors are
 1-$\sigma$). 

Terrestrial Gamma-Ray Flashes turn out  to be  very efficient
particle accelerators in our atmosphere. Our detected power-law emission
between 10 MeV and 100 MeV is difficult to reconcile  with
current RREA  models
\cite{rousseldupre,lehtinen,Dwyer2005,Dwyer2008,carlson1,gurevich2}.
Some of these models are characterized by acceleration over typical distances near, e.g.,
 stepped-leader lightning sizes ($\sim 50-100$ m) that correspond to a small number of
avalanche lengths.
On the contrary, an observed photon energy of 100 MeV implies a lower limit on the
acceleration distance $d_{min}    \simeq  (1\, {\rm km}) \, (\bar{E}_{100})^{-1}$ where
$\bar{E}_{100}$ is the average electric field in units of 100 kV/m. These large-scale sizes
are a significant fraction of the intra-cloud or cloud-to-ground distances over which
potential drops of order of 100 MV can be established in thunderstorms.
Furthermore, the detection of TGF emission in
the 10-100 MeV range renews the interest for the neutron production in
these energetic events as well as in normal lightning. A deteiled
report on this discovery is reported in \cite{Tavani2010}.

\subsection{Gamma-ray localization of TGFs}

Since the discovery that TGFs can exhibit energies well above 20~MeV,
we started searching for TGF events in the GRID, the AGILE gamma-ray
imager sensitive above 20~MeV. The added value provided by the GRID is
due to its imaging capabilities so that, for the first time, a TGF
could be not only detected, but also localized from space directly in gamma-rays. 

In the period between June~2008 and December~2009 the MCAL
instrument triggered 119 bursts identified as TGFs according to
the selection criteria discussed in \cite{Marisaldi2010}. For each
of these bursts, the GRID  dataset was searched for
quasi-simultaneous gamma-ray events within a 200~ms time-window
centered at the TGF start time $T_0$, defined as the time of the first  MCAL-photon associated with the TGF. 
A peak in the cumulative distribution of the arrival times is
evident for the 2~ms time bin immediately following $T_0$. This
peak includes 13 events, and the probability for it to be a
statistical fluctuation (13 events or higher) is $6.5^. 10^{-10}$  if we assume
that GRID events are not correlated to TGFs and are distributed
according to the Poisson law with the measured average rate of
5.1~counts/s. 
 All these GRID events take place during the TGF emission time
 interval estimated from MCAL data only.

\begin{figure}
\includegraphics{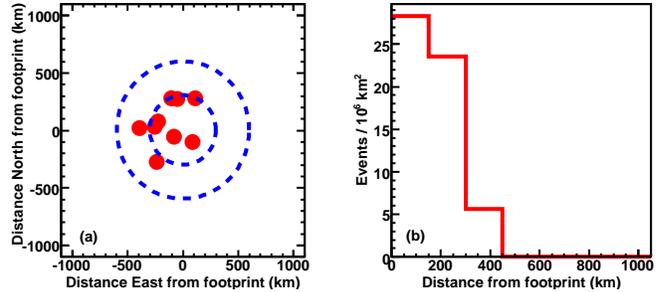}     
\caption{\label{fig2} (a) scatterplot of the GRID events projection
  with respect to the AGILE footprint. The dashed circles are 300 and
  600~km in radius. The full AGILE visibility region has a 2600~km
  radius. (b) occurrence density vs. distance from footprint (each bin
  has been divided by the subtended area in $\mathrm{km}^2$). Figure
  from \cite{Marisaldi2010b}.}
\end{figure}

Figure \ref{fig2} shows the scatterplot of the GRID events projection with respect to the AGILE footprint (the point at the Earth's surface on the straight line joining the satellite and the Earth's centre), and the distribution of the occurrence density vs. distance from footprint (each bin has been divided by the subtended area in $\mathrm{km}^2$). 
The GRID photon directions are obtained by the standard analysis pipeline for photons within the AGILE field of view, and by a custom version of the same software for photons outside the field of view. The photon directions are then back-projected from the satellite position, obtained by GPS data, to the Earth's surface, parameterized as the World Geodetic System WGS-84 ellipsoid, in order to obtain the production site location. 
All 9 events are contained within a 1.14~sr solid angle, a factor 3.4  smaller than the solid angle subtended by the Earth at
the satellite altitude of 540~km,  which corresponds to a maximum visibility projected distance radius of $\sim 2600$~km from the satellite footprint. If the GRID events originate  directly from the TGF production site, these results are consistent with a distance to footprint less than $\sim 300$~km estimated for RHESSI TGFs using sferics data \cite{Cummer2005,Hazelton2009,Cohen2010,Connaughton2010}.

The AGILE Tracker detections, described in details in
\cite{Marisaldi2010b}, provide indeed the
first direct TGF localization from space.
The GRID detection of TGFs above 20~MeV in about 7\% of the cases confirms that TGFs emit a
substantial amount of energy above this value, as reported in the
previous section.

\subsection{TGF correlation with global lightning activity}

Correlation of TGFs with lightning activity was established shortly
after their discovery. Several detailed one-to-one correlations of
TGFs with individual lightning strokes have been published, see
\cite{Connaughton2010,Lu2010} for some recent results, but a global
climatology of TGFs is still missing, although some steps
towards this direction have been reported \cite{Smith2010,Splitt2010}.

\begin{figure*}
\includegraphics[width=150mm]{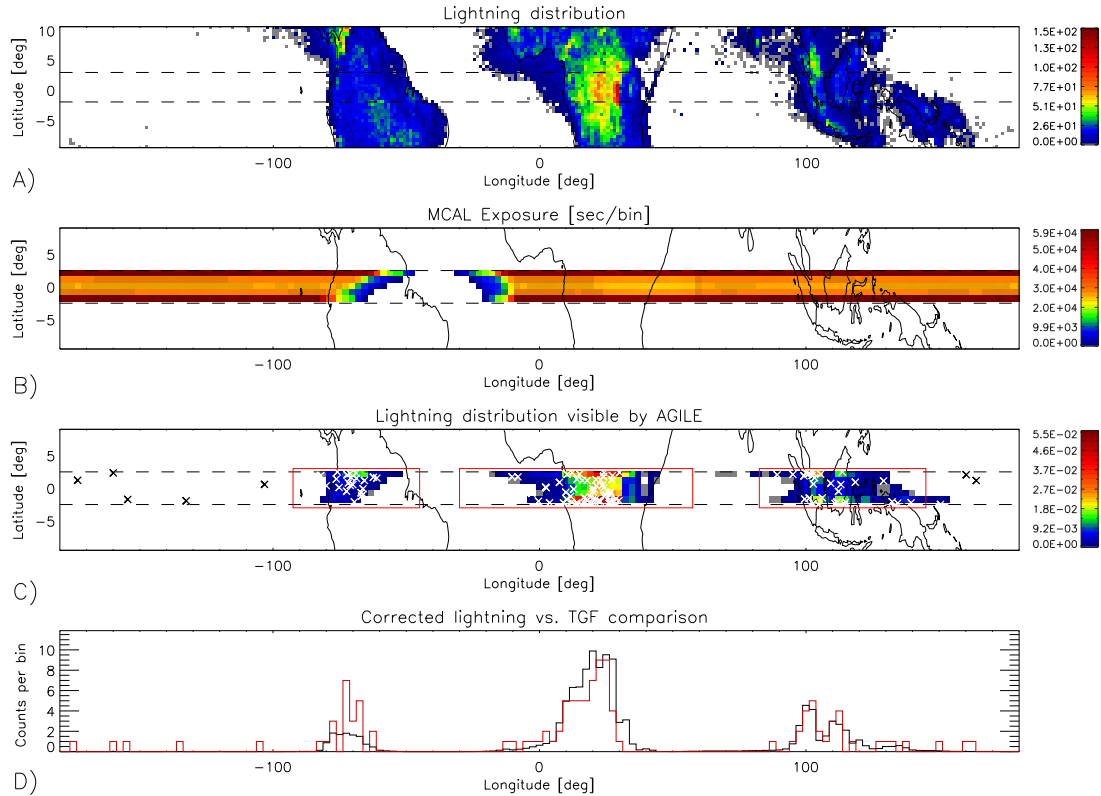}
\caption{Lightning and TGF maps.
A: LIS-OTD high resolution full climatology flashes rate [flashes/km$^2$/year] (0.5x0.5 deg per bin);
B: MCAL exposure map [sec/bin] (2.5x1.0 deg per bin);
C: LIS-ODT multiplied by MCAL exposure [flashes/km$^2$] (2.5x1.0 deg per bin). The crosses indicate the 
AGILE-TGF locations.
The red borders indicate the continental zones considered in the analysis; 
D: Longitude distributions, summed over all latitudes, of the AGILE-TGF map (red) and LIS/OTD corrected map 
(black), corresponding to the maps showed in panel \ref{global}-C,
normalized to the total number of TGFs. Figure
  from \cite{Fuschino2011}
}\label{global}
\end{figure*}

Thanks to its detection capabilities and its very low inclination orbit, AGILE guarantees a high exposure
above a narrow equatorial belt. Indeed, with about 120 TGFs/year, the AGILE satellite record 
the highest TGF  detection rate surface density. This capability allowed
a detailed high spatial resolution study of the correlation between
TGFs and global lightning activity as detected from space by the
Optical Transient Detector (OTD), onboard the Orbview-1 spacecraft,
and the Lightning Imaging Sensor (LIS), onboard the TRMM NASA mission,
(LIS/OTD public data available at http://ghrc.msfc.nasa.gov/). The aim
of this correlation study, based on mono and bi-dimensional
Kolmogorov-Smirnov test, is to link TGFs to average properties of
lightning geographical distribution.
Figure \ref{global}-A shows the annual lightning flash rate density per km$^2$ per year with a spatial resolution of 0.5 degrees for 
both longitude and latitude, based on LIS/OTD High Resolution Full Climatology 
(HRFC) data comprising about 10 years of observations (1995-2005)
\cite{Christian2003}. Figure \ref{global}-B shows the MCAL exposure map,
accounting also for the periods during which the trigger logic was
active, while figure \ref{global}-C shows the global lightning distribution multiplied by the MCAL exposure, that is
directly comparable with AGILE TGF distribution. The bin size of 2.5 deg in longitude and 1 deg in latitude 
corresponds, on ground, to about 275 km and 110 km respectively. In the same image all AGILE TGF positions are 
also shown, with white and black crosses.
Finally, in Figure \ref{global}-D the TGF (red) and the exposure-corrected lightning (black) longitude 
distributions are shown, summed over all latitudes.

At a qualitative level, the longitudinal distributions for TGF and lightning above the continental areas (Figure \ref{global}-D) 
show quite good agreement, and it is possible to recognize the main
features of the three continental areas, like the sharp cut over the
Congo or the double-peaked feature due to Sumatra and Borneo
islands. However, to assess the compliance between the TGF and
lightning distributions in a quantitative way, we used the
Kolmogorov-Smirnov test (KS hereafter), which provides a probability value (P) for the null hypothesis that two unbinned data sets are drawn from the same distribution.
After correction for the AGILE exposure, we find that we cannot
consider the global TGF distribution as a random sample of the
lightning distribution. Moreover, we find significant regional
differences in the degree of correlation. In particular, in the case
of south east Asia we find a 87\% probability for the TGF distribution
being a sub-sample of lightning, while this probability surprisingly
drops to 3\% in the case of Africa. 

The results shown here are an independent confirmation of those
reported in the work based on RHESSI data \cite{Smith2010} concerning
an excess of TGFs above central America and south east Asia and a
corresponding depletion above Africa. However, reference
\cite{Smith2010} reports an eastward shift between TGF and lightning
distribution above south east Asia, which is currently not supported
by the the very good agreement we found for the two distributions
above that region, but we must note that our result is obtained over a
much narrower latitude region than that considered in that work. 
Moreover we must consider that the TGF dataset considered span very
different periods of time. In fact, while AGILE dataset includes 118
TGFs detected in 12 months, the $1^{st}$ RHESSI TGF catalog
\cite{Grefenstette2009} includes 144 TGFs in the same latitude range,
but detected during 102 months of observation. So, while RHESSI TGFs
are up to now the most complete sample for climatological studies at a
global scale, AGILE exhibits the higher detection rate surface density
(TGFs/month per square degree) over a narrower geographical region
limited by its orbital inclination, allowing detailed studies of this
climatic region. 

Based on the crude assumption that the observed TGF/lightning ratio
holds at all latitudes we can also estimate a global rate of $\dot{N}
\simeq 220 \div 570$ TGFs per day, in agreement with previous
estimates \cite{Carlson2009}. 
The correlation study presented in this section is described in
details in reference \cite{Fuschino2011}.

\section{CONCLUSIONS} 

AGILE is succesfully observing TGFs since June 2008 and is currently one of the only three operating space instruments capable of detecting TGFs. The AGILE payload is very well suited for TGF science. Its main strength points can be summarized as follows:

\begin{itemize}
\item[-] AGILE-MCAL effective area peaks in the MeV range, the range where most of the TGF energy is radiated;
\item[-] the MCAL energy range is extended up to 100~MeV, allowing to probe the high energy tail of the TGF spectrum;
\item[-] the trigger logic on time scales as short as 1ms and 290$\mu s$, well matching the TGF typical time scale, makes the AGILE sample not biased toward the brightest/longest events;
\item[-] the MCAL design strategy, spatial segmentation in several independent detection units, makes the instrument less sensitive to dead-time and pile-up effects than monolithic detectors of equivalent volume;
\item[-] event data with $2\mu s$ timing accuracy are available for triggered events: time binning is limited by counting statistics only;
\item[-] absolute timing accuracy better than $100 \mu s$ allows precise timing for correlation with on-ground observation of sferic waves associated to lightnings;
\item[-] the AGILE-GRID trigger logic is sufficiently flexible to collect also high-energy photons coming from the Earth;
\item[-] the AGILE orbit with $2.5^\circ$ inclination is optimal for mapping the equatorial region, where most of the TGFs take place, with exposure much larger than other missions;
\end{itemize}

Thanks to these capabilities, several important results on TGF science
based on AGILE observations have already been published
\cite{Marisaldi2010,Tavani2010,Marisaldi2010b,Fuschino2011},
concerning the yet poorly understood TGF high-energy component, the first 
localization of TGFs from space, and the correlation of TGFs with
global lightning activity.

\begin{acknowledgments}
AGILE is a mission of the Italian Space Agency (ASI), with co-participation of INAF
(Istituto Nazionale di Astrofisica) and INFN (Istituto Nazionale di Fisica Nucleare).
Research partially funded through the ASI contract n. I/089/06/2.

The v2.2 gridded satellite lightning data were produced by the NASA LIS/OTD Science Team (Principal
Investigator, Dr. Hugh J. Christian, NASA / Marshall Space Flight Center) and are available from the Global
Hydrology Resource Center (http://ghrc.msfc.nasa.gov).
\end{acknowledgments}

\bigskip 

\end{document}